\documentclass[%
 reprint,
 amsmath,amssymb,
 aps,
]{revtex4-1}

\usepackage{graphicx}% Include figure files
\usepackage{dcolumn}% Align table columns on decimal point
\usepackage{bm}% bold math
\begin{document}

\preprint{APS/123-QED}

\title{A Discussion on Possible Effects of the Barbero-Immirzi Parameter at
the TeV-scale Particle Physics}% Force line breaks with \\
%\thanks{A footnote to the article title}%

\author{N. Panza}
\email{npanza@cefet-rj.br}

\author{H. Rodrigues}
\email{harg@cefet-rj.br}
\affiliation{Departamento de F\'{\i}sica, Centro Federal de Educa\c{c}\~{a}o Tecnol\'{o}gica Celso Suckow da
Fonseca \\
Av. Maracan\~{a}, 229, 20271-110, Rio de Janeiro, RJ, Brazil}%

\author{D. Cocuroci}
\email{cocuroci@cbpf.br}

\author{J. A. Helay\"{e}l-Neto}
\email{helayel@cbpf.br}
\affiliation{Centro Brasileiro de Pesquisas F\'{\i}sicas \\
Rua Dr. Xavier Sigaud 150, 22290-180, Rio de Janeiro, RJ, Brazil}% \textbackslash\textbackslash}

% This line break forced with \textbackslash\textbackslash
%}%

\date{\today}% It is always \today, today,
             %  but any date may be explicitly specified

\begin{abstract}
In this paper, we analyse a curvature- and torsion-square quantum gravity action with an additional Holst term minimally coupled to a
massive Dirac field in four dimensions. The main purpose here is to try to estimate and compare
the value of the Barbero-Immirzi (BI) parameter with its currently known results. To do that, we work out the physical mass of the fermion as a function of this parameter in a perturbative one-loop calculation, assuming the scenario of a physics at the TeV-scale. 
\begin{description}
%\item[Usage]
%Secondary publications and information retrieval purposes.
\item[PACS numbers]
12.10.-g, 12.38.Bx, 04.50.kd.
%\item[Structure]
%You may use the \texttt{description} environment to structure your abstract;
%use the optional argument of the \verb+\item+ command to give the category of each item. 
\end{description}
\end{abstract}

\pacs{Valid PACS appear here}% PACS, the Physics and Astronomy
                             % Classification Scheme.
%\keywords{Suggested keywords}%Use showkeys class option if keyword
                              %display desired
\maketitle

%\tableofcontents

\section{Introduction}\label{sec:intro}

It is well-known that the Einstein-Cartan theory represents the simplest
extension of General Relativity (GR) in the presence of the torsion field \citep{Freidel2005,Peixoto2012,Felisola2013}.
However, it is a non-renormalizable quantum theory, in spite
of describing physics with a very good approximation at the classical level. In view of that, several
investigations followed in order to set up approaches to the gravitational
interaction that may prove relevant to our understanding of the quantum gravity
problem. A fruitful approach has been the theory of higher-order curvature
invariants, whose actions contain terms quadratic in the curvature,
considering also all possible quadratic terms that we can form with torsion, in
addition to the Einstein-Hilbert term. Furthermore, these terms are necessary
if we wish to obtain an effective action for quantum gravity at scales close
to the Planck length \citep{Capozziello2007,Capozziello2009}. This proposal is not yet free from problems.
The main interest in such theories comes, in particular, from Cosmology. In
this case, the possibility that the graviton have a small mass to explain the
early stages of the Universe depends on the particular quantum theory of gravity. Moreover,
torsion could, in principle, be associated to any astrophysical scale.

In a more contemporary context of non-perturbative approaches to quantum
gravity, Loop Quantum Gravity arose as another possible quantization method \citep{Ashtekar1986,Ashtekar1991,Gambini1996,Rovelli2004,Thiemann2007}. 
The heartwood of this scheme is the idea that the fundamental
variable for quantization is the connection, rather than the metric. It was
proposed by Asthtekar \citep{Ashtekar1986,Ashtekar1991} by working only with the self-dual part of the
Hilbert-Palatini action. In this formulation, the phase space variable is a
complex SU(2)-connection. Unfortunately, this approach exhibits a difficult step,
namely, to recover the phase space of real general relativity, we must impose
reality conditions on the original complex Asthtekar formulation of gravity.
In order to come over this difficult, Barbero and, later on Immirzi introduced a
family of canonical variables, associated to the BI parameter \citep{Barbero1995,Immirzi1997} (henceforward 
represented by $\beta$) which is actually a  canonical theory of
gravity in terms of a real SU(2)-connection, referred to as Asthtekar-Barbero
connection. In a Lagrangian approach, the $\beta$- parameter is taken as a new
dimensionless parameter introduced with the meaning of a coupling constant of a topological 
odd-parity term added to the Hilbert-Palatini action, which is treated in the first-order
formalism and yields the so-called Holst action \citep{Holst1996}. This topological
term, in GR, does not affect the equations of motion in the absence of torsion
since it appears as a factor in a term that vanishes on the mass-shell.
Although $\beta$ is non-physical at classical level, we find that it becomes an
essential parameter at the quantum level, appearing in several contexts, such as,
the spectrum of area and volume operators \citep{Rovelli1996}, in the black hole entropy
formula \citep{Wang2014} and in the dynamics provided by the more modern spin foam models
\citep{Dittrich2012,Freidel2008}. It should be noticed that there is another method to quantize gravity
known as the quantum Regge calculus \citep{Rocek1981,Rocek1984}, which is $\beta$-independent.

For these reasons, over the latest years, several attempts to fix the $\beta$-
parameter have been worked out. Dreyer \citep{Dreyer2003}, motivated by Hod's work \citep{Hod1998} on the
asymptotic quasi-normal modes of black holes, proposed a novel way to fix this
parameter. In order that the Bekenstein-Hawking entropy formula and
quasi-normal mode agree, the BI parameter should
be given by $\beta=\ln3/\left(  2\sqrt{2}\pi\right)  $. In a recent paper,
considering Faddeev's formulation of gravity, Khatsymovsky \citep{Khatsymovsky2012} found out that the
elementary area spectrum is proportional to the BI parameter
with the value $\beta=0,39$. Requiring the validity of the Bekenstein-Hawking
area law relating the two parameters that characterize the quantum states of
black hole horizon, A. Majhi et al
obtained the following range for $\beta$: $0.159<\beta<0.225$ \citep{Majhi2012,Majhi2013}. In the
latest paper that, to our knowledge, addresses this issue, the authors have obtained the value
$\beta=\ln3/\pi$, through the calculation of the entropy for an arbitrary non-rotating
isolated horizon \citep{Wang2014}.

It is worthy mentioning that all references previously quoted estimate the $\beta$-
parameter making use of a non-perturbative approach to QG. In this paper, we
follow an alternative route, by studying perturbative QG \citep{Benedetti} in the first-order
(tetrad) formalism, to understand which is the role of the $\beta$- parameter in
the loop corrections to a model discussed in Ref. \citep{Hernaski2010}, enriched by
implementing the Holst term, with minimal coupling between fermions and
torsion. This choice is motivated by the fact that, in the Einstein-Cartan theory,
the torsion field becomes relevant only in the presence of fermionic currents
and, more recently, coupling fermions to classical GR has attracted a great deal of
attention, since that A. Perez and C. Rovelli \citep{Perez2005} have demonstrated that the minimal
coupling to fermions renders gravity sensitive to the $\beta$- parameter \citep{Shapiro2014}. We
tackle the issue of a 1-loop mass generation mechanism for the fermion as well as
the possible influence of the BI parameter on this mechanism.

The outline of this work is the following. In Section \ref{sec:2}, we briefly
review the main results of the Holst action at the Lagrangian level and we
present a description of the model discussed in Ref. \citep{Hernaski2010}, as well as our
conventions together with the propagators of the model. Also, since the Dirac
field couples to the affine connection and the intrinsic spin of fermions acts
as a source for torsion, in Section \ref{sec:3}, we study an extension of the our
model in order to include the minimal coupling between torsion and the Dirac
spinor. Finally, in Section \ref{sec:4}, we cast our Discussions and Conclusions.

\section{Description of the model}\label{sec:2}

We adopt to work with the first-order formalism, in which  configuration space consists
of two independent field variables, namely, the space-time $SO\left(  4\right)
$ gauge spin connection, $\omega_{\mu}$ $^{ab\text{ }}$, and the vielbein,
$e_{\mu}$ $^{a}$ . This choice is justified for we think that it is a more
fundamental approach to gravitation, since it is based on the fundamental
ideas of the Yang-Mills approach. The  gravitational Lagrangian
(Einstein-Cartan term) together with the Holst invariant, where the latter
is multiplied by a coupling constant $\beta\in \mathbb{R} -\left\{0\right\}$, the BI parameter, can be written as follows:
\begin{equation}%
\mathcal{L}_{EC}+\mathcal{L}_{H}=e\left(-\alpha R+\frac{\alpha}{2\beta}e_{a}^{\mu}e_{b}^{\nu}%
\epsilon^{ab}\text{ }_{cd}R_{\mu\nu}\text{ }^{cd}\right) , \label{1}
\end{equation}
where $\alpha=\frac{1}{16\pi G}$, $e$ denotes the absolute value of the
determinant of the co-tetrad , $R_{\mu\nu}$ $^{cd}$ is the field strength
associated with the spin connection $\omega_{\mu}$ $^{ab}$ which is not
torsion-free, and $\epsilon_{abcd}$ denotes the 4-dimensional Levi-Civita
tensor. In pure gravity, $\beta$ does not affect the graviton equation of
motion. This is no longer the case when fermions are present. In this respect,
the effective Dirac action contains an axial current-current interaction 
whose dependence on $\beta$ becomes non-trivial; also, this parameter cannot take an
arbitrary value, so that we have to fix it by calculating some observable effect.

Our motivation in this paper is to investigate the possible relevance of the BI
parameter at the quantum level, considering the most general parity-preserving
Lagrangian that describe massive gravitons studied in the Ref. \citep{Hernaski2010} with an
additional Holst term, as given below:
\begin{eqnarray}
\mathcal{L}&=& e\left( -\alpha R  +\chi R^{2} + \rho R_{\mu a}R^{\mu a} + \gamma R_{\mu a}R^{a\mu} \right. \nonumber \\
&& + \left.    \xi R_{\mu\nu ab}R^{\mu\nu ab}+kR_{\mu\nu ab}R^{ab\mu\nu}  + \lambda R_{\mu\nu ab}R^{\mu a\nu b}\right. \nonumber \\ 
&& + \left.  xT_{\mu ab}T^{\mu ab}+yT_{\mu ab}T^{ab\mu}+zT_{\mu a}\text{ }^{a}T^{\mu b}\text{ }_{b}  \right. \nonumber \\            
&& + \left.   \frac{\alpha}{2\beta}e_{a}^{\mu}e_{b}^{\nu}\epsilon^{ab}\text{ }%
_{cd}R_{\mu\nu}\text{ }^{cd}\right) ,  \label{2}
\end{eqnarray}
where $\chi$, $\beta$, $\gamma$, $\xi$, $k$ , $\lambda$ and $\beta$ are
arbitrary dimensionless constants and the others parameters have canonical
dimensions given by $\left[  s\right]  =\left[  t\right]  =$ $\left[
r\right]  =1$.

The particular choice of the Lagrangian above, Eq. (\ref{2}), is justified by the fact that, by suitably choosing special regions in parameter space, it yields a healthy spectrum of excitations, with massive gravitons and no ghosts or tachyons present, despite the presence of higher powers of curvature and torsion terms. This has been discussed in the works of Refs. \citep{Hernaski2010,Helayel2010}.  Other terms with powers of torsion could be considered; they would however spoil the consistency of the action, in that ghosts or tachyons would be present and would not decouple from the physical sector \citep{Hernaski2010,Helayel2010}. This is why we restrict ourselves to the terms present in the action (\ref{2}).  

Our conventions are:
\begin{eqnarray}
R_{\mu\nu}\text{ }^{ab}&=&\partial_{\mu}\omega_{\nu}\text{ }^{ab}-\partial_{\nu
}\omega_{\mu}\text{ }^{ab}+\omega_{\mu}\text{ }^{a}\text{ }_{c}\omega_{\nu
}\text{ }^{cb} \nonumber \\
&& - \, \omega_{\nu}\text{ }^{a}\text{ }_{c}\omega_{\mu}\text{ }^{cb} , \label{3}
\end{eqnarray}%
\begin{equation}
R_{\mu}\text{ }^{a}=e_{b}^{\nu}R_{\mu\nu}\text{ }^{ab} , \label{4}
\end{equation}
\begin{equation}
R=e_{a}^{\mu}e_{b}^{\nu}R_{\mu\nu}\text{ }^{ab}, \label{5}
\end{equation}
and
\begin{equation}
\eta_{\mu\nu}=\left( 1,-1,-1,-1\right) , \label{6}
\end{equation}
where the Greek indices refer to the world manifold and the Latin ones stand
for the frame indices.

 The torsion tensor has 24 independent components in four space-time dimensions and, by considering the irreducible representations of $SO(1,3)$, it may be split into a vector, $v^{\mu}$, an axial vector $a^{\mu}$, and a
rank-three tensor $q_{\mu\nu\alpha}$, which satisfies the the conditions
$q_{\mu\alpha}$ $^{\alpha}=0$ and $\epsilon^{\alpha\beta\mu\nu}q_{\alpha
\beta\mu}=0$. With this in mind, the generic torsion can be expressed as \citep{Shapiro2014}
\begin{equation}
T_{\nu\mu\alpha}=q_{\nu\mu\alpha}+\frac{1}{3}\left(  \eta_{\nu\alpha}v_{\mu
}-\eta_{\mu\alpha}v_{\nu}\right)  +\epsilon_{\nu\mu\alpha\lambda}a^{\lambda} , \label{7}
\end{equation}
where $g_{\mu\nu}=\eta_{ab}e_{\mu}$ $^{a}e_{\nu}$ $^{b}$ is the metric
tensor. The splitting of Eq. (\ref{7}) for $T_{\nu\mu\alpha}$ in $SO(1,3)$-irreducible components is simply a group-theoretic decomposition. So long as a dynamical model is concerned, it is perfectly legitimate to truncate some of the components in RHS of Eq. (\ref{7}) if we choose to restrict our considerations to a Cartan's restricted geometry, which amounts to taking $q_{\nu\mu\alpha}=0$. Moreover, this irreducible piece and the $v^{\mu}$-vector component of the torsion do not couple to the fundamental fermions of the Standard Model. This is the reason why we allowed to suppress them and we build up our model in such a way that only the $a^{\mu}$-vector acquires dynamical and can be excited by the interaction with the fundamental fermions of the Standard Model.

Next, we perform the split of the vielbein,
\begin{equation}
e^{\mu}\text{ }_{a}=\delta^{\mu}\text{ }_{a}+h^{\mu}\text{ }_{a} , \label {8}
\end{equation}
namely, the Minkowski background and a perturbation, $h^{\mu}$ $_{a}$, around the background. In addition, the
fluctuation can be decomposed as
\begin{equation}
h_{ab}=\frac{1}{4}\eta_{ab}s+\epsilon_{abcd}\partial^{c}W^{d} . \label{9}%
\end{equation}

Before we proceed, we recall that the spin connection $\omega^{\mu}$ $_{ab}$
may be expressed as%
\begin{equation}
\omega_{\mu}\text{ }^{ab}=\widetilde{\omega}_{\mu}\text{ }^{ab}+K_{\mu}\text{
}^{ab} , \label{10}%
\end{equation}
where $\widetilde{\omega}_{\mu}$ $^{ab}$ is the Riemannian part of the spin connection
\begin{equation}
\widetilde{\omega}_{\mu}\text{ }^{ab}=\frac{1}{2}e_{\mu c}\left(  \Omega
^{cab}+\Omega^{acb}-\Omega^{bac}\right) , \label{11}%
\end{equation}
where $\Omega_{cba}=e_{c}^{\mu}e_{b}^{\nu}\left(  \partial_{\mu}e_{\nu
a}-\partial_{\nu}e_{\mu a}\right)$ stands for the rotation coefficients.
$K_{\mu}\text{ }^{ab}$ are the components of the contorsion tensor, defined by
\begin{equation}
K_{\mu}\text{ }^{ab}=\frac{1}{2}\left( T_{\mu}\text{ }^{ab}-T^{ab}\text{ }_{\mu
}+T^{b}\text{ }_{\mu a}\right) . \label{12}
\end{equation}
It is worthy noticing that, while the contortion tensor is antisymmetric in the
last two indices, the torsion tensor is antisymmetric in the first two indices.

Using equations (\ref{7}) to (\ref{12}), the spin connection fluctuation may be put into a
more useful form, namely:
\begin{eqnarray}
\omega_{\mu}\text{ }^{\alpha\beta}&=&\frac{1}{8}\left( \delta_{\mu}\text{
}^{\beta}\partial^{\alpha}s-\delta_{\mu}\text{ }^{\alpha}\partial^{\beta}s\right) \nonumber \\ 
&& + \frac{1}{2}\left(  \epsilon^{\beta}\text{ }_{\mu\rho\lambda
}\partial^{\alpha}\partial^{\rho}W^{\lambda}  - \epsilon^{\alpha}\text{ }_{\mu\rho\lambda}\partial^{\beta} 
\partial^{\rho}W^{\lambda}\right) \nonumber \\
&&  + \frac{1}{3}\left(  \delta_{\mu}\text{
}^{\beta}v^{\alpha}-\delta_{\mu}\text{ }^{\alpha}v^{\beta}\right)  + 
\frac{1}{2}\epsilon_{\mu}\text{ }^{\alpha\beta}\text{ }_{\lambda}a^{\lambda} . \label{13}
\end{eqnarray}

At this point, it is worthy to stress that in the presence of torsion, the Gauss-Bonnet term does not correspond any longer to a topological invariant, for it cannot be expressed as a total divergence \citep{Niu2008}. Thus, all the quadratic terms in the curvature present in the Lagrangian (\ref{2}) must be kept in our considerations.

Finally, we can obtain the relevant part of the bilinear expansion of the
Lagrangian (\ref{2}). Making use of equation (\ref{13}), after a lengthy but
straightforward calculation, we obtain
\begin{eqnarray}
\mathcal{L}&=& s\left[ \frac{3}{16}\left(  \rho+\gamma+\xi+3\chi+k+\frac{\lambda}{2}\right)
\square-\frac{39\alpha}{32}\right]  \square s  \nonumber \\
&& + v^{\mu}\left[  \frac{7\alpha}{2}-\left(  \frac{7\rho}{6}+\gamma+\xi-3\chi
+k+\frac{5\lambda}{12}\right)  \square\right]  \partial_{\mu}s  \nonumber \\
&& + v^{\mu}\left[ -\left(  \frac{\rho}{4}+\frac{2\lambda}{9}+\frac{8\xi}%
{9}\right)  \eta_{\mu\nu}\square \right.  \nonumber \\
&& \left. - \frac{4}{3}\left(  \frac{2\rho}{3}%
+\gamma-\frac{\xi}{3}+k+3\chi+\frac{\lambda}{3}\right)  \partial_{\mu}\partial_{\nu
}\right. \nonumber \\
&&  \left.  +  \left(  \frac{2\alpha}{3}+\frac{2x}{3}-\frac{y}{3}+z\right)  \eta
_{\mu\nu}\right]  v^{\nu} \nonumber \\
&& + w^{\mu}\left[  \left(  \frac{\rho}{2}-\frac{\gamma
}{2}+2\xi-2k\right)  \eta_{\mu\nu}\square^{2}  \right. \nonumber \\
&& \left. +  \left(  -\frac{\rho}{2}-\frac{\gamma}{2}-2\xi+2k-\lambda\right)
\square\partial_{\mu}\partial_{\nu} \right.  \nonumber \\
&& \left. - \frac{5\alpha}{2}\eta_{\mu\nu}%
\square+\frac{5\alpha}{2}\partial_{\mu}\partial_{\nu}\right]  \square w^{\nu} \nonumber \\
&& + a^{\mu}\left[  \left(  \rho-\gamma+4\xi-4k-\frac{\lambda}{2}\right)
\square^{2}+2\alpha\eta_{\mu\nu}\square  \right. \nonumber \\
&& \left. + 2\alpha\partial_{\mu}\partial_{\nu
} + \left( -\rho+\gamma-4\xi+4k-\frac{5\lambda}{2}\right)  \square
\partial_{\mu}\partial\nu\right]  w^{\nu}  \nonumber 
\end{eqnarray}
\begin{eqnarray}
&& + a^{\mu}\left[  \left(  \frac{\rho}{2}-\frac{\gamma}{2}+2\xi-2k\right)
\square \right.  \nonumber \\
&& \left. + \left(  -\frac{\rho}{2}+\frac{\gamma}{2}+\xi+2k-\frac{3\lambda}%
{2}\right)  \partial_{\mu}\partial_{\nu} \right. \nonumber \\
&& \left. - 3\left(  \frac{\alpha}{2}+2x+2y\right)  \eta_{\mu\nu}\right]  a^{\nu} \nonumber \\
&& -  a^{\mu}\left(  \frac{21\alpha}{8\beta}\right)  \partial_{\mu}s-a^{\mu
}\left(  \frac{2\alpha}{\beta}\right)  \eta_{\mu\nu}v^{\nu} . \label{15}
\end{eqnarray}

Two important observations are in order. The $w^{\mu}$-field is not sensitive to the
BI parameter, so that we decide to truncate it from the decomposition of the metric.
The second observation concerns the fact that, in the next Section, we shall
consider the minimal fermion coupling in the presence of massive Dirac fields.
As well-known, a Dirac particle only interacts with the totally antisymmetric
part of the torsion, so that we can disregard the vector component of the
torsion. Therefore, the Lagrangian (\ref{15}) can be written only in terms of the
scalar component of the metric and the pseudo-vector component of the
torsion. We understand that only the $s$- and $a^{\mu}$-
fields should contribute effects whenever fermions are introduced. This is why
we truncate the $v^{\mu}$- and $w^{\mu}$- fields from the Lagrangian (\ref{15}). Thus,
\begin{eqnarray}
\mathcal{L}&=&\frac{3}{16}\left(  \rho+\gamma+\xi+k+3\chi+\frac{\lambda}{2}\right)  s\square
^{2}s-\frac{39\alpha}{32}s\square s  \nonumber \\
&& + \left(  -\frac{\rho}{2}+\frac{\gamma}{2}-2\xi+2k\right)  \left(  \partial
_{\mu}a^{\nu}\right)^{2}  \nonumber \\
&& + \left(  \frac{\rho}{2}-\frac{\gamma}{2}%
-\xi-2k+\frac{3\lambda}{2}\right)  \left(  \partial_{\mu}a^{\mu}\right)^{2} \nonumber \\
&& - 3\left(  \frac{\alpha}{2}+2x+2y\right)  a^{\mu}a_{\mu}-\frac{21\alpha}{4\beta
}a^{\mu}\partial_{\mu}s . \label{16}
\end{eqnarray}

The coefficients in the expression (\ref{16}), except the constant $\alpha$ are
arbitrary parameters. By setting
\begin{equation}
-\frac{\rho}{2}+\frac{\gamma}{2}-2\xi+2k=0 , \label{17a}
\end{equation}%
\begin{equation}
\frac{\rho}{2}-\frac{\gamma}{2}-\xi-2k+\frac{3\lambda}{2}=0 , \label{17b}
\end{equation}
and
\begin{equation}
x+y=0,  \label{17c}
\end{equation}
we get $\xi=\lambda/2$. So, the Lagrangian (\ref{16}) can be written in terms of the new
parameters as
\begin{equation}
\mathcal{L} = As\square^{2}s+Fs\square s+Ba^{\mu}a_{\mu}+Ha^{\mu}\partial_{\mu}s , \label{18}
\end{equation}
where
\begin{equation}
A=\frac{3}{16}\left(  \rho+\gamma+\xi+k+3\chi+\frac{\lambda}{2}\right)  \ ,
\ \ \ \ B=-\frac{3\alpha}{2} , \label{19a}%
\end{equation}
\begin{equation}
F=-\frac{39\alpha}{32}  , \ \ \ \ \ \ \ \ \ \ \ \ \ \ H=-\frac{21\alpha
}{4\beta} . \label{19b}
\end{equation}

If we make use of equation (\ref{18}), we can immediately obtain the equation of motion for the
axial vector field, $a^{\mu}$, which reads
\begin{equation}
a_{\mu}= - \frac{7}{4\beta}\partial_{\mu}s  . \label{20}%
\end{equation}

With the free parameters of our model chosen according to the relations of Eqs. (\ref{17a})-(\ref{17c}), we are led to the Lagrangian of Eq. (\ref{18}), where it is clear that the vector $a^{\mu}$ becomes an auxiliary field: it appears algebraically, so it has no independent dynamics. Therefore, it can be solved by its algebraic equation of motion in terms of the $s$ - field, as it will be done below.

For the sake of a better comprehension, we should perhaps justify why the relations of Eqs. (\ref{17a})-(\ref{17c}) are imposed. The reason behind the choices of Eqs. (\ref{17a}) and -(\ref{17b}) is to suppress the unphysical (ghost-type) excitation that would show up from $\left(\partial_{\mu}a^{\mu}\right)$ and from the symmetric part $\left(\partial_{\mu}a^{\nu}\right)$: the spin - 0 ghost carried by $\left(\partial_{\mu}a^{\mu}\right)$ and $\left(  \partial_{\mu}a^{\nu}+\partial_{\nu}a^{\mu}\right)  $ would spoil the tree-level unitarity of our model. Finally, condition (\ref{17c}) is assumed in order to exclude from our model a Planckian mass that would otherwise be present in the spectrum of excitations of $a^{\mu}$, as the calculation of its propagator reveals.

Hence, the Lagrangian (\ref{18}) finally takes the form
\begin{equation}
\mathcal{L} = As\square^{2}s+\left(F+\frac{H^{2}}{4B}\right)  s\square s  . \label{21}%
\end{equation}

Before going on, one remark is in order. The next step is to read off the
Feynman propagator for the scalar field $s$. After a direct computation from
(\ref{21}), we attain%
\begin{equation}
D_{F}\left(  p\right)  =\frac{i}{2\left(  F+\frac{H^{2}}{4B}\right)  }\left\{
-\frac{1}{p^{2}}+\frac{1}{p^{2}-m^{2}}\right\} , \label{22}%
\end{equation}
where
\begin{equation}
m^{2}=\frac{1}{A}\left(  F+\frac{H^{2}}{4B}\right) . \label{23}%
\end{equation}

Since that we are working in scenarios with low scales, up to a few Tev, and
the mass $m$ is at the Planck scale, the Appelquist-Carazzone theorem
ensures that the effects of the Planckian loops are not detectable in the Tev
scale. So, in view of the suppression of the effects of the Planckian massive
excitation ($p^{2}=m^{2}$) by powers of $1/m^{2}$, for the sake of our
considerations the propagator for the scalar field may be taken as follows:
\begin{equation}
D_{F}\left(  p\right)  =-\frac{i}{2\left(  F+\frac{H^{2}}{4B}\right)  p^{2}}  .
\label{24}
\end{equation}

Once we have analyzed the spectrum of excitations that arise from the gravity
sector, we shall be now concerned with the introduction of fermions and our
main goal shall be the computation of the effects of the (virtual) gravity
models on the fermion self-energy and, in doing so, we shall track the
effects the BI parameter contributes to the fermion masses.

\section{Coupling with fermionic matter fields}\label{sec:3} 

Over the past years, there has been a great deal of interest in the study of Dirac fields in
curved space-time. Particularly, it was shown by Perez and Rovelli [26] and by
Freidel, Minic and Takeuchi [1] that, whenever minimally coupled, spinor fields 
present in the Einstein-Cartan theory with the Holst term prevent the latter
of being a truly topological action. In presence of fermions, we have a
modification in the structure of the space-time, leading to a non-vanishing
torsion field. Therefore, we have a classical-level interpretation of this
parameter, which is related with axial-axial current interactions. Now, in
order to better understand the effect of the BI parameter in Particle
Physics, let us calculate the corrections to the fermion mass due to this
parameter in the model describe in the previous Section. 

The Lagrangian density which corresponds to the Dirac equation in curved
space-time is
\begin{equation}
\mathcal{L}_{D}=e\left(  \frac{i}{2}e^{\mu}\text{ }_{a}\overline{\psi}\gamma^{a}D_{\mu
}\psi-\frac{i}{2}e^{\mu}\text{ }_{a}D_{\mu}\overline{\psi}\gamma^{a}\psi
-m_{0}\overline{\psi}\psi\right) . \label{24}%
\end{equation}

The covariant derivative acting on a spinor is%
\begin{equation}
D_{\mu}\psi=\partial_{\mu}\psi+\frac{g}{8}\omega_{\mu}\text{ }^{ab}\left[
\gamma_{a}, \gamma_{b}\right]  \psi  ,\label{25}%
\end{equation}
where $\gamma_{\mu}=e_{\mu}$ $^{a}\gamma_{a}$ , $a$, $\mu=0$, $1$, $2$, $4$,
$\gamma_{a}$ are the usual flat space-time Dirac matrices, $g$ is a
dimensionless coupling constant and $m_{0}$ is the mass of the Dirac field.
According to the discussion in the Section \ref{sec:2}, we are working only with the
scalar field, $s$, and the pseudo-vector, $a^{\mu}$, so that the spin connection can
be written as%
\begin{equation}
\omega_{\mu}\text{ }^{\alpha\beta}=\frac{1}{8}\left(  \delta_{\mu}\text{ }^{\beta
}\partial^{\alpha}s-\delta_{\mu}\text{ }^{\alpha}\partial^{\beta}s\right)
+\frac{1}{2}\epsilon_{\mu}\text{ }^{\alpha\beta}\text{ }_{\lambda}a^{\lambda}  .
\label{26}%
\end{equation}

Replacing (\ref{25}) and (\ref{26}) into (\ref{24}), and making use of the identity
\begin{equation}
\gamma^{a}\gamma^{b}\gamma^{c}=\gamma^{a}\eta^{bc}+\gamma^{c}\eta^{ab}%
-\gamma^{b}\eta^{ab}+i\epsilon^{abcd}\gamma_{5}\gamma_{d} , \label{27}%
\end{equation}
we arrive, after some algebra, to the relation
\begin{eqnarray}
\mathcal{L}_{D}&=&\frac{i}{2}\left(  \overline{\psi}\gamma^{\mu}\partial_{\mu}\psi
-\partial_{\mu}\overline{\psi}\gamma^{\mu}\psi\right)  -m_{0}\overline{\psi
}\psi \nonumber \\
&& + \frac{is}{2}\left(  \overline{\psi}\gamma^{\mu}\partial_{\mu}\psi
-\partial_{\mu}\overline{\psi}\gamma^{\mu}\psi\right)  +\frac{3g}{4}(1+s)a^{\mu
}\overline{\psi}\gamma_{5}\gamma_{\mu}\psi  \nonumber  \\
&& -m_{0}%
s\overline{\psi}\psi  . \label{28}
\end{eqnarray}

Using equation (\ref{21}), together the Dirac Lagrangian (\ref{21}), we can obtain the new equation of motion for the auxiliary axial vector field, $a^{\mu}$, which reads%
\begin{equation}
a^{\mu}=-\frac{7}{4\beta}\partial^{\mu} s + \frac{g}{4\alpha}(1+s)  \overline{\psi} \gamma_{5}\gamma^{\mu}\psi  . \label{29a}
\end{equation}
Here, it is worthwhile to recall that, since $a^{\mu}$ appears as an auxiliary field, it is perfectly licit procedure to replace it in the Dirac coupled action by its equation of motion (\ref{29a}). Previously, $a^{\mu}$ was given by Eq. (\ref{20}). With the coupling to the fermions, the axial vector condensate $\overline{\psi}\gamma^{\mu}\gamma_{5}\psi$ contributes and, as it follows below, upon replacement of $a^{\mu}$ by Eq. (\ref{29a}) in the original Dirac action, one obtains the following interaction Lagrangian: 
\begin{eqnarray}
\mathcal{L}_{int}&=&\frac{is}{2}\left(  \overline{\psi}\gamma^{\mu}\partial_{\mu}%
\psi-\partial_{\mu}\overline{\psi}\gamma^{\mu}\psi\right)  -\frac{21g}{16\beta}(1+s)\partial_{\mu
}s\overline{\psi}\gamma_{5}\gamma^{\mu}\psi  \nonumber \\
&& +\frac{3g^2}{16\alpha}(1+s)^2(\overline{\psi}\gamma_{5}\gamma^{\mu}\psi)^2-m_{0}%
s\overline{\psi}\psi  ,  \label{29}
\end{eqnarray}
where now a 4-fermion contact appears as a consequence of the non-dynamical character of the $a^{\mu}$-axial field. 

One can observe that the coupling constant of this interaction depends on the BI parameter.
Let us consider the Feynman rules for the Lagrangian (\ref{21}). The momenta in this
paper are incoming to the vertices. Thus, the fermion-scalar 3-vertex takes
the form
\begin{equation}
\frac{i}{2}\left(  p+p\right)  _{\mu}\gamma^{\mu}-\frac{21g}{16\beta}q_{\mu}\gamma_{5}%
\gamma^{\mu}-im_{0} , \label{30}
\end{equation}
and the other two vertices that describes the four-fermion interaction are given by%
\begin{equation}
\frac{21g}{16\beta}q_{\mu}\gamma_{5}\gamma^{\mu} , \label{31}
\end{equation}
and
\begin{equation}
\frac{3g^2}{16\alpha}\gamma_{5}\gamma^{\mu}\gamma_{5}\gamma_{\mu}  .
\end{equation}

The propagator for the spinor field is given by
\begin{equation}
S_{F}\left(  p\right)  =\frac{i}{p\!\!\!/-m_{0}} . \label{32}
\end{equation}

From these results, we turn into the calculation of the self-energy corrections
to the fermion propagator and discuss the mechanism of mass generation for
the fermion field, by looking at the pole(s) of  its 1-loop corrected 2-point function
For the fermion self-energy graph, we find
%\begin{equation}
%-i\Sigma\left(  p\right)  =%
%{\displaystyle\int}
%\frac{d^{4}q}{\left(  2\pi\right)  ^{4}}\left\{  \frac{i}{2}\left(
%2p\!\!\!/-q\!\!\!/\right)  -\frac{21g}{16\beta}\gamma_{5}q\!\!\!/ -im_{0}\right\}   \label{33}%
%\end{equation}%
%\[
% \times i\frac{p\!\!\!/-q\!\!\!/+m_{0}}{\left(  p-q\right)  ^{2}-m_{0}^{2}}i\left(
%-\frac{1}{2\widetilde{F}q^{2}}\right)  \left\{  \frac{i}{2}\left(
%2p\!\!\!/-q\!\!\!/\right)  -\frac{21g}{16\beta}\gamma_{5}q\!\!\!/-im_{0}\right\} ,
%\]

\begin{eqnarray}
-i\Sigma\left(  p\right) & =& \int \frac{d^{4}q}{\left(  2\pi\right)  ^{4}}\left\{  \frac{i}{2}\left(
2p\!\!\!/-q\!\!\!/\right)  -\frac{21g}{16\beta}\gamma_{5}q\!\!\!/ -i m_{0}\right\}   \nonumber \\
&&  \times i\frac{p\!\!\!/-q\!\!\!/+m_{0}}{\left(  p-q\right)  ^{2}-m_{0}^{2}}i  
\dfrac{16\beta
^{2}}{3\alpha q^{2}\left( 13\beta ^{2}+49\right) }  \nonumber \\
&& \times  \left\{  \frac{i}{2}\left(
2p\!\!\!/-q\!\!\!/\right)  -\frac{21g}{16\beta}\gamma_{5}q\!\!\!/-im_{0}\right\} \nonumber \\
&& - Tr \int \frac{d^{4}k}{\left( 2\pi \right) ^{4}}\dfrac{3g^{2}}{16\alpha 
}i\gamma _{5}\gamma ^{\mu }\gamma _{5}\gamma _{\mu }i\dfrac{k\!\!\!/+m_{0}}{%
k^{2}-m_{0}^{2}} .  \label{33}%
\end{eqnarray}

The result for the divergent part of the 1-loop self-energy of the fermion
has the following final form%
%\begin{eqnarray}
%\widetilde{\Sigma}\left(  p\right)  &=&\frac{1}{16\pi^{2} \varepsilon \widetilde{F}
%}\left\{  m_{0}\left[  m_{0}^{2}\left(  \frac{1}{4}+a^{2}\right)
%-3p^{2}\right]  I_{4\times4}\right. \nonumber \\  
%&& \left. + \frac{1}{2}\left[  \frac{1}{3}\left(  \frac{79}{4}-7a^{2}\right)
%p^{2}+\left(  \frac{1}{4}-a^{2}\right)  m_{0}^{2}\right]  p_{\mu}%
%\gamma^{\mu} \right.   \nonumber \\
%&& \left. +  \frac{ia}{2}\left(  \frac{19}{3}p^{2}+m_{0}^{2}\right)  p_{\mu}%
%\gamma^{\mu}\gamma_{5}\right\} ,  \label{38}
%\end{eqnarray}

%\begin{eqnarray}
%\widetilde{\Sigma}\left(  p\right)  &=&  \frac{1}{\pi^{2} \alpha \varepsilon } 
%\left\{  \left\{ \frac{2 \beta^2}{3(13 \beta^2 + 49)}  \left[ 3 \beta^2 - \frac{m_0^2}{4} \left( 1 + \frac{441 g^2}{64 \beta^2} \right)
%\right] \right.  \right. \nonumber \\
%&& \left. - \frac{3 g^2 m_0^2}{8} \right\}  I_{4\times4}  \nonumber \\
%&& + \frac{\beta^2}{12(13 \beta^2 + 49)} \left[\frac{1}{3}\left(79-\frac{3087 g^2}{64\beta^2} \right)\beta^2 \right.  \nonumber \\  
%&& \left. + \left(\1-\frac{441 g^2}{64 \beta^2} \right) m_0^2 \right]  p_{\mu} \gamma^{\mu} \nonumber \\
%&& + \frac{7 \i g \beta}{48 \left( 13 \beta^2 +49 \right)}  \left(  19 \beta^2 + 3 m_0^2 \right) p_{\mu}%
%\gamma^{\mu}\gamma_{5} \right\} , \label{38}
%\end{eqnarray}

\begin{eqnarray}
\widetilde{\Sigma}\left(  p\right)  &=&\frac{1}{\pi ^{2}\alpha \varepsilon }\left\{ \left\{ \frac{2\beta
^{2}}{3\left( 13\beta ^{2}+49\right) }\left[ p ^{2}-\frac{m_{0}^{2}}{4}%
\left( 1+\frac{441g^{2}}{64\beta ^{2}}\right) \right] \right. \right.  \nonumber \\
&&\left. -\frac{3g^{2}m_{0}^{2}}{8}\right\} I_{4\times 4} \nonumber \\
&&+\frac{\beta ^{2}}{12\left( 13\beta ^{2}+49\right) }\left[ \frac{1}{3}%
\left( 79-\frac{3087g^{2}}{64\beta ^{2}}\right) p ^{2}\right.  \nonumber \\
&&+\left. \left( 1-\frac{441g^{2}}{64\beta ^{2}}\right) m_{0}^{2}\right]
p_{\mu }\gamma ^{\mu } \nonumber \\
&&+\left. \frac{7 i g\beta }{48\left( 13\beta ^{2}+49\right) }\left(
19p ^{2}+3m_{0}^{2}\right) p_{\mu }\gamma ^{\mu }\gamma _{5}\right\} , \label{38}
\end{eqnarray}
where $\varepsilon^{-1}%
=\ln\left(  \Lambda^{2}/\mu^{2}\right)  $ is the parameter of dimensional
regularization$,\Lambda$ is the cut-off and $\mu$ is an arbitrary mass parameter.

The insertion of this 1-loop graph leads to the following 1-loop corrected
Lagrangian%
\begin{equation}
\mathcal{L}=\overline{\psi}\left(  \gamma^{\mu}p_{\mu}-m_{0}-\widetilde{\Sigma}\right)
\psi=\overline{\psi}O\psi   .  \label{39}
\end{equation}

We are now ready to find the 1-loop corrected fermionic propagator and then the
relation between the bare and the physical mass of the fermion in our
model. The practical calculation for that follows the standard scheme and we do not
report the details. The result is the on-shell relation
\begin{eqnarray}
&& \left(D^{2}-E^{2}+C^{2}-2E\cdot C\right)  \left(  D^{2}-E^{2}+C^{2}+2E\cdot
C\right) \nonumber \\
&&  + \, 4  E^{2}C^{2} = 0 , \label{40}
\end{eqnarray}
where we have used the simplified notation, $E^{2}=E_{\mu}E^{\mu}$, and the fact that
$p^{2}=m^{2}$. We omit the full result of eq. (\ref{40}), which is very lengthy. We
report below only the parts of interest for our actual considerations:
\begin{widetext}
\begin{equation}
D=-m_{0}\left\{  \frac{2\beta^{2}}{3\pi^{2}\alpha\epsilon\left(  13\beta
^{2}+49\right)  }\left[  3m^{2}-\frac{1}{4}\left(  1+\frac{441g^{2}}%
{64\beta^{2}}\right)  m_{0}^{2}\right]  +1\right\} ,  \label{41}
\end{equation}
\end{widetext}
\begin{widetext}
\begin{equation}
E_{\mu}=\left\{  1+\frac{\beta^{2}}{12\pi^{2}\alpha\epsilon\left(  13\beta
^{2}+49\right)  }\left[  \frac{1}{3}\left(  79-\frac{3087g^2}{64\beta^{2}%
}\right)  m^{2}+\left(  1-\frac{441g^{2}}{64\beta^{2}}\right)  m_{0}%
^{2}\right]  \right\}  p_{\mu}  , \label{42}
\end{equation}%
\end{widetext}
\begin{equation}
C_{\mu}=\frac{7ig\beta}{48\pi^{2}\alpha\epsilon\left(  13\beta
^{2}+49\right)  }\left(  19m^{2}+3m_{0}^{2}\right)  p_{\mu} .  \label{43}
\end{equation}

We could not achieve an analytic expression for the mass, $m$, as
function of the $\beta$ parameter from Eq. (\ref{40}). Therefore, the remainder of
our calculation will be carried out using MAPLE computer algebra system. In the
next Section, we present the results of our numerical analysis of the model we are
discussing and compare them with the known results for the BI parameter cited in
the Introduction.

\section{Discussions and Conclusions}\label{sec:4}

\begin{figure}[htpb] %[htpb]
\centering 
\includegraphics[width=.30\textheight]{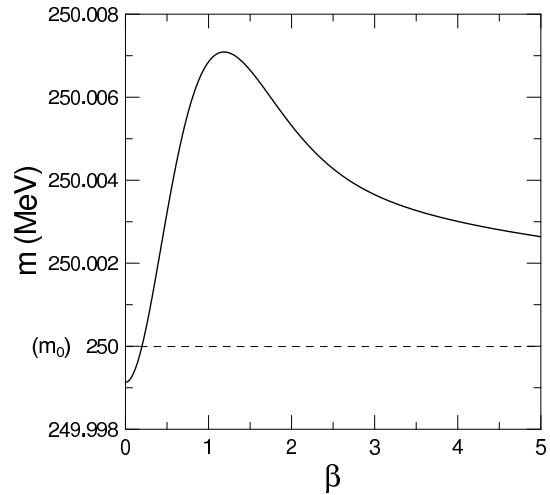}
\vspace*{-.1cm} %8pt}
\caption{The running of $\beta$, with $\Lambda= 250$ Mev, $\mu= 1$ Mev,
$m_{0} = 250$ Mev.  \label{fig:100}}
\end{figure}

%\begin{figure}[htpb] %[htpb]
%\centering 
%\includegraphics[width=.30\textheight]{Figure1.eps}
%\vspace*{-.1cm} %8pt}
%\caption{The running of $\beta$, with $\Lambda= 250$ Mev, $\mu= 1$ Mev,
%$m_{0} = 250$ Mev.  \label{fig:100}}
%\end{figure}

%\begin{figure}[htpb] %[htpb]
%\centering 
%\includegraphics[width=.30\textheight]{Figure2.eps}
%\vspace*{-.1cm} %8pt}
%\caption{The running of $\beta$, depicted in the previous figure, in the interval $0.159 \leq \beta \leq  0.225$. Notice
%the almost linear behavior. \label{fig:200}}
%\end{figure}

Our main interest in this paper has been the study of the effect of the BI parameter on
the mass of a fermion in the context of a particular Extended Theory of
Gravity with explicit torsion terms. For this purpose, we have performed one-loop
calculations in this model with gravity and fermions minimally coupled.
Proceeding this way, and adopting the physical mass of the fermion at the LHC scale, we expected to obtain an estimate for
the BI parameter. Unfortunately, this could not be done, for the equation
(\ref{40}) depends on the free parameters, namely, the cut-off $\Lambda$, the
arbitrary mass parameter $\mu$ and the bare mass of the fermion, $m_{0}$. For
this reason, we proceed further to adopt an alternative route. As a first
step, we fix the parameters $\Lambda$, $\mu$ and $m_{0}$, choosing an
appropriate energy scale, such as the LHC energy range. With these considerations in mind,
we obtain a relation between $m$ and $\beta$. Thus, one can check if the range
of $\beta$, which has currently been reported in the literature, is compatible with the
observed mass of the fermion in the energy scale chosen. In the Figure \ref{fig:100}, we cast the results obtained for $\Lambda=250$ MeV (corresponding to QCD cut-off),
$\mu=1$ MeV, $g=1$, $m_{0}=250$ MeV and $0<\beta<5.0$. As one should expect,
the corrections to the fermion mass are extremely small. By inspection of the plot, 
we see that the fermion mass increase if the BI
parameter lies in the range $0<\beta < $1.185, reaching its maximum value at $\beta$ = 1.185. In this interval of values, the potential term in Dirac interaction Lagrangian dominates the kinetic term. Furthermore, when $\beta\rightarrow0$, the fermion mass is finite. In particular, in the range $0<\beta<0.195$, the physical mass of the fermion is less than his bare mass. On the other hand, for values of $\beta$ greater than 1.185, the kinetic term dominates the potential term. In conclusion, our result is
compatible with the known results in the literature for the BI parameter.

The point of view we have tried to convey in this paper is that the present established limits for the BI parameter are compatible with the spectrum of fermions with masses at the LHC scale. Our study shows that the Standard-Model charged leptons and quarks would not be sensitive to the BI parameter, but we understand that the whole sector of TeV-scale massive fermions, such as the charginos and neutralinos, could be a good probe of the BI parameter in the realm of Particle Physics. We would like to point out that the model we present here is still too limited. To our mind, the next immediate step towards a more realistic model, namely, the inclusion of the BI parameter in the framework of Supersymmetry (more specifically, the minimal SUGRA model), would be the right landscape for the investigation we are reporting in this paper. We are now  concentrating efforts in this direction and we expect to report soon our first efforts elsewhere.

\begin{acknowledgments}
D. Cocuroci and J. A. Helay\"{e}l-Neto express their appreciation to CNPq - Brazil and FAPERJ - Rio de Janeiro for the invaluable financial support. The authors are also grateful to Ant\^{o}nio Duarte for his pertinent suggestions on original manuscript of this paper.
\end{acknowledgments}

\end{document}